\documentclass[USletter,10pt]{article}
\usepackage{geometry}               
\usepackage{epstopdf}
\usepackage{amsthm}
\usepackage{amsfonts}
\usepackage{MnSymbol}
\usepackage{hyperref}
\usepackage{color}
\usepackage{tikz}
\usepackage{graphicx}
\usepackage[affil-it]{authblk}
\usepackage[applemac]{inputenc} 
\usepackage[toc,page]{appendix}

\theoremstyle{plain}
\newtheorem{thm}{Theorem}[section]
\newtheorem{lem}[thm]{Lemma}
\newtheorem{prop}[thm]{Proposition}

\theoremstyle{definition}
\newtheorem{defn}{Definition}[section]

\theoremstyle{remark}
\newtheorem*{rem}{Remark}

\newcommand{\bea}{\begin{eqnarray}}
\newcommand{\eea}{\end{eqnarray}}
\newcommand{\RN}[1]{%
  \textup{\uppercase\expandafter{\romannumeral#1}}%
}
\def\beaa{\begin{eqnarray*}}
\def\eeaa{\end{eqnarray*}}
\def\ba{\begin{array}}
\def\ea{\end{array}}
\def\be#1{\begin{equation} \label{#1}}
\def \eeq{\end{equation}}
\def\bsplit{\begin{split}}

\begin{document}
\title{On the local extension of Killing vector fields in electrovacuum spacetimes}
\author{Elena Giorgi} 
\maketitle
\begin{abstract}
   We revisit the problem of extension of a Killing vector field in a spacetime which is solution to the Einstein-Maxwell equation. This extension has been proved to be unique in the case of a Killing vector field which is normal to a bifurcate horizon in \cite{PinYu}. Here we generalize the extension of the vector field to a strong null convex domain in an electrovacuum spacetime, inspired by the same technique used in \cite{ExtensionRicciFlat} in the setting of Ricci flat manifolds. We also prove a result concerning non-extendibility: we show that one can find local, stationary electrovacuum extension of a Kerr-Newman solution in a full neighborhood of a point of the horizon (that is not on the bifurcation sphere) which admits no extension of the Hawking vector field. This generalizes the construction in \cite{ExtensionRicciFlat} to the electrovacuum case.
\end{abstract}

\section{Introduction}
In (pseudo)-Riemannian manifolds, Killing vector fields are generators of local isometries. They therefore play a major role in the description of the symmetries of the manifold. In addition, in Mathematical General Relativity the identification of Killing vector fields in spacetimes veryfing the Einstein equation is closely related to the no-hair theorem for black holes. Neverthless, the problem of extension of Killing vector fields in (pseudo)-Riemannian manifolds can be stated independently of its applications to General relativity (see Section \ref{relation-rigidity} of Introduction for a summary of the relation with the problem of rigidity of black holes).

The main purpose of the present paper is to address some aspects of the following problem. Suppose that a pseudo-Riemannian manifold possesses a local isometry in some open subsets of the manifold. Then, it is a natural question to ask if such symmetry can be extended past the open subset, or if the open subset is the maximal set on which the isometry holds. The problem can more specifically be formulated as follows.

 Suppose $(M,g)$ is a pseudo-Riemannian manifold, and let $Z$ be a Killing vector field defined on an open subset $O \subset M$. We are interested in identifying the necessary conditions on the manifold and on the subset $O$ for which $Z$ can be uniquely extended to be a Killing vector field in $(M,g)$. The problem has been previously studied and solved in the following cases:
 \begin{itemize}
 \item If $(M,g)$ is an analytic and simply connected manifold, a classical result of Nomizu (\cite{Nomizu}) assures the uniqueness of the extension of a Killing field defined on a connected open domain $O$. 
 
 In this case, the condition of analyticity is crucial to prove the existence of the desired extension. Indeed, in \cite{Nomizu}, the author proves that in an analytic (pseudo)-Riemannanian manifold, at every point the dimension of the germs of Killing vector fields is constant in a neighborhood of the point. Therefore, together with the assumption of simply connectedness, the prolongation of the Killing vector field obtained as solution of a second order ordinary differential equation can be extended to the whole manifold.

 \item If $(M,g)$ is a smooth Ricci flat manifold and $O \subset M$ is bounded by a bifurcate horizon, then the local extension of a Killing vector field normal to the horizon is unique in a full neighborhood, as proved by Alexakis, Ionescu and Klainerman in \cite{Hawking's}. Their work extends a previous result of Friedrich, R\'acz and Wald, where the prolongation of the Killing vector field was limited to the domain of dependence of the bifurcate horizon (and therefore not applicable to the problem of rigidity of black holes). 
 
 In \cite{Hawking's}, the strong structure of the boundary, which is a bifurcate regular non-expanding horizon, implies the existence of a Killing vector field (known as the Hawking Killing vector field) along the null generators of the horizon. Such Killing vector field is then proved to be extendible as a Killing field in a full neighborhood of the horizon. As in Nomizu's result, the Killing vector field is extended as a solution to an ordinary differential equation. More precisely, the vector field $Z$ on the horizon is extended along a geodesic transversal to the horizon and generated by a vector field $L$ using the commutator relation $[Z,L]=L$. To prove that such extension is still a Killing vector field in the case of a smooth (and not analytic) manifold, the authors make use of Carleman estimates to prove unique continuation.

 \item If $(M,g)$ is a smooth Ricci flat manifold and $O \subset M$ is a strongly null convex domain\footnote{See Definition \ref{definingfunction} in Section \ref{unique-continuation}.} at a point $p \in \partial O$, then the local extension is unique in a neighborhood of the point $p$ in $M$, as proved by Ionescu and Klainerman in \cite{ExtensionRicciFlat}.

 In their work, as opposed to the case of the bifurcate horizon above, the domain $O$ does not possess a particular structure which implies the existence of a Killing vector field along its boundary. Therefore the existence of a Killing vector field up to the boundary of $O$ is assumed, as well as a convexity condition (which was implied in the previous case by the structure of the bifurcate horizon). In \cite{ExtensionRicciFlat}, the Killing vector field is extended through the Jacobi equation. This allows the authors to use the hypothesis of Ricci flatness to derive a system of transport and covariant wave equations to which Carleman estimates are applied to prove a unique continuation result.
  
 \item If $(M,g)$ is a smooth manifold verifying the Einstein-Maxwell equation and $O \subset M$ is bounded by a bifurcate horizon, then the local extension of a Killing vector field normal to the horizon is unique in a full neighborhood of the horizon, as proved by Yu in \cite{PinYu}. His work extends the result in \cite{Hawking's} from solutions to the Einstein vacuum equation to solutions to the Einstein-Maxwell equation.

The proof of the extension in the case of electrovacuum spacetimes follows the same steps as in \cite{Hawking's}: the vector field extended through the commutator relation is proved to be Killing by Carleman estimates. 
 \end{itemize}
 
The aim of this paper is to prove the extension of a Killing vector field defined in a strongly null-convex domain in an electrovacuum spacetime. We state here the theorem.

\begin{thm}\label{Theorem} Assume that $(M,g)$ is a smooth manifold which verifies the Einstein-Maxwell equation, and let $O \subset M$ be a strongly null convex domain at a point $p \in \partial O$. Suppose that the metric $g$ admits a smooth Killing vector field $Z$ in $O$ that preserves the electromagnetic tensor in the Einstein-Maxwell equation. Then $Z$ extends to a Killing vector field for $g$ to a neighborhood of the point $p$ in $M$. Moreover, this extension preserves the electromagnetic tensor. 
\end{thm}

The proof of the theorem relies on the extension of the Killing vector field through the Jacobi equation, as inspired by \cite{ExtensionRicciFlat}. The main difficulty in the generalization of the result in \cite{ExtensionRicciFlat} to the case of electrovacuum spacetime is the coupling of the curvature with the electromagnetic tensor given by the Einstein-Maxwell equation. This results in a system of transport and covariant wave equations to which adapted Carleman estimates and unique continuation techniques are applied.

As a generalization of the result proved in \cite{PinYu}, Theorem \ref{Theorem} applies to any pseudo-Riemannian manifolds of any dimension verifying the Einstein-Maxwell equation, for vector fields $Z$ not necessarily tangent to $\partial O$ in a neighborhood of $p$.

\subsection{Relation to the black hole rigidity problem in General Relativity}\label{relation-rigidity}

The results in this paper are motivated by the rigidity problem in General Relativity, or the so called no-hair theorem. 

The theorem goes back to Hawking (see \cite{Haw}), who proved that real analytic stationary solutions of the Einstein field equations possessing a horizon are isometric to Kerr black holes. Hawking's proof roughly goes as follows. The stationary Killing vector field on the horizon implies the existence of an axially symmetric Killing vector field on it. The assumption of analyticity of the manifold then allows to extend the axially symmetric Killing vector field past the event horizon by a Nomizu-type argument. Finally, the classification of stationary and axisymmetric solutions given by the Carter-Robinson theorem then implies the rigidity of the solution to be a Kerr black hole.

It is therefore clear that the extension of Killing vector fields in manifolds verifying the Einstein equation, possibly coupled with matter fields, is very relevant to the black hole rigidity problem. In the case of solutions to the Einstein-Maxwell equation, a rigidity theorem of stationary and axially symmetric solutions is given by the Mazur-Bunting's theorem, making a similar argument given by Hawking possible in the case of Kerr-Newman black holes.

Theorem \ref{Theorem} represents a step towards the resolution of the rigidity problem of charged black holes in the more realistic sets of smooth manifolds. In particular, the result of Theorem \ref{Theorem} should be thought as applied to the axially symmetric Killing vector field of a charged black hole at the bifurcation sphere of its horizon. Its local extension, as proved in this paper, then allows to apply the Mazur-Bunting's theorem and deduce a local isometry to the Kerr-Newman black hole.

\

As required in Theorem \ref{Theorem}, the condition of strong null convexity is needed in order to obtain the desired extension. For this reason, in the setting of rigidity of charged black holes the extension should be done at the bifurcation sphere of the horizon, where the convexity condition is verified. In particular, in any other point of the horizon the result fails to be true.

In the second theorem proved in this paper, we provide a counterexample to extendibility.
 In \cite{ExtensionRicciFlat}, the authors provided a counterexample in Kerr spacetime to the extension of the axially symmetric Killing vector field from a point that is not on the bifurcation sphere (see Theorem 1.3 in \cite{ExtensionRicciFlat}).
  Here we generalize the counterexample to the Kerr-Newman spacetime $\mathcal{K}_{m,a,Q}$. In Theorem \ref{Counterexample}, we show that one can modify the Kerr-Newman spacetime smoothly, on one side of the event horizon of the black hole, in such a way that the resulting metric is still an electrovacuum spacetime, has a stationary Killing field $T=\partial_{t}$, but does not admit an extension of the axially symmetric Killing vector field.

\ 

The paper is organized as follows. 

In section 2 we present few preliminaries to electrovacuum spacetimes. In section 3 we prove the unique extension result (Theorem \ref{Theorem}) and in section 4 we present the counterexample to the extendibility (Theorem \ref{Counterexample}). In Appendix A we derive the formalism of Ernst potential of a Killing vector field used in the construction of the counterexample. 

\

\textbf{Acknowledgements} The author would like to thank Sergiu Klainerman and Mu-Tao Wang for suggesting the problem and for valuable discussions.

\section{Preliminaries}

Throughout this paper, let $M$ be a connected smooth pseudo-Riemannian manifold.

A pseudo-Riemannian $n$-dimensional manifold $(M,g)$ is called an \textit{electrovacuum spacetime} if the metric $g$ satisfies the Einstein-Maxwell equation 
 \bea
\label{Einsteineq}
\operatorname{Ric}-\frac 1 2 R g=\mathcal{T}(F)
\eea
 where $\operatorname{Ric}$ is the Ricci tensor of $(M,g)$, $R$ its scalar curvature and $\mathcal{T}(F)$ the stress-energy tensor associated to an electromagnetic tensor $F$, given in coordinates by 
 \bea\label{def-stress-energy}
 \mathcal{T}(F)_{ij}=2g^{lk}F_{il}F_{jk}-\frac 1 2 g_{ij} |F|^2
 \eea
 where $|F|^2=g^{ij}g^{kl}F_{ik}F_{jl}$.
  The electromagnetic tensor $F$ is a $2$-form that satisfies the Maxwell equations, i.e. 
 \bea
 \label{divergenceF}
 D^iF_{ij}=0.
 \eea
 \bea
 \label{BianchiF}
 D_i F_{jk}+D_j F_{ki}+ D_k F_{ij}=0
 \eea
 where $D$ denotes the Levi-Civita connection of $(M,g)$. 
 \begin{rem}By construction, the stress-energy tensor is trace free. By taking the trace with respect to $g$ of \eqref{Einsteineq} we see that the scalar curvature tensor of an electrovacuum spacetime always vanishes. Therefrre the Einstein equation reduces to $\operatorname{Ric}=\mathcal{T}(F)$.
 \end{rem}
 
 We summarize in the following proposition some properties of the Ricci curvature and of the electromagnetic tensor of an electrovacuum spacetime.\footnote{Note that we denote with $R$ the curvature of the manifold, and the indices will clarify if we are referring to the Riemann tensor or the Ricci tensor.  When we prefer an expression without indices we will denote them $\operatorname{Riem}$ and $\operatorname{Ric}$ respectively.}
 \begin{prop}\label{boring-relations} In an electrovacuum spacetime, the following relations hold true:
\bea
D^i R_{ijkl}&=&D_k R_{jl}-D_l R_{jk} \label{divergenceRiemann} \\
\Box_g F_{jk}&=& \left({{R^i}_{jk}}^m-{{R^i}_{kj}}^m\right)F_{im}+{{R}_{j}}^mF_{mk}-{{R}_{k}}^mF_{mj} \label{waveF}\\
\Box_g R_{ij}&=& 2g^{lk}\left( F_{jk}\Box_g F_{il}+F_{il}\Box_g F_{jk}+ 2 D_mF_{il}D^mF_{jk}\right) - \left( {F_l}^k \Box_g {F_k}^l+ 2 |D F|^2\right) g_{ij} \label{waveRicci}
\eea
\bea
\label{waveRiemann}
\begin{split}
\Box_g R_{jklm}&= D_j(D_lR_{km}-D_mR_{kl})-D_k(D_lR_{jm}-D_mR_{jl}) +{R_j}^p R_{kpml} -{R_k}^pR_{pjlm} \\
&+{{R^i}_{jk}}^p R_{iplm}-{{R^i}_{kj}}^pR_{iplm} +{{R^i}_{jl}}^pR_{kimp}-{{R^i}_{jm}}^pR_{kilp}+{{R^i}_{kl}}^pR_{ijmp}-{{R^i}_{km}}^pR_{ijlp}
\end{split}
\eea
\bea
\label{waveDF}
\begin{split}
\Box_g D_m F_{kj}&=D_m \left(({{R^i}_{jk}}^l -{{R^i}_{kj}}^l) F_{li} + {R_j}^l F_{kl} - {R_k}^l F_{jl}\right)+ D^i{{R_{mi}}^l}_j F_{kl} -D^i {{R_{mi}}^l}_k F_{jl} \\
&+{{R_{mi}}^l}_j D^iF_{kl}- {{R_{mi}}^l}_k D^i F_{jl}+{{R^i}_{mj}}^l D_l F_{ki}-{{R^i}_{mk}}^l D_l F_{ji}+({{R^i}_{mk}}^l D_j- {{R^i}_{mj}}^l D_k)F_{li}+{R_m}^l D_l F_{kj}
\end{split}
\eea
where $\Box_g= D^k D_k$ is the D'Alembertian operator of $(M,g)$.
 \end{prop}
 \begin{proof} Equation \eqref{divergenceRiemann} is a direct consequence of the second Bianchi identity. 
 
 The wave equation for $F$ is a consequence of Maxwell equation \eqref{BianchiF}. In fact, applying $D^i$ to the equation and commuting covariant derivatives,
 \beaa
0&=& D^i D_i F_{jk}+D^iD_j F_{ki}+ D^iD_k F_{ij}\\
 &=&\Box_g F_{jk} +D_jD^iF_{ki}+{{R^i}_{jk}}^mF_{mi}+{{R^i}_{ji}}^mF_{km}+D_kD^iF_{ij}+{{R^i}_{ki}}^mF_{mj}+{{R^i}_{kj}}^mF_{im}
 \eeaa
 and using equation (\ref{divergenceF}) to eliminate the divergence of $F$, we get the desired expression.
 
 The wave equation for $\operatorname{Ric}$ is a consequence of the Einstein equation \eqref{Einsteineq}.  
 The wave equation for $\operatorname{Riem}$ is a conseguence of the second Bianchi identity and equation (\ref{divergenceRiemann}). Indeed, differentiating and contracting the second Bianchi identity we get 
 $$D^i D_i R_{jklm}+D^i D_j R_{kilm}+D^i D_k R_{ijlm}=0$$ Commuting the derivatives in the second and third term and substituting the divergence with equation (\ref{divergenceRiemann}), we obtain the desired expression.
 
 The wave equation for $DF$ is a consequence of Maxwell equations \eqref{divergenceF} and \eqref{BianchiF}. In fact, applying $D_m$ to the equation (\ref{BianchiF}) and commuting derivatives:
 \beaa
 0=D_m D_i F_{jk}+D_m D_j F_{ki}+ D_m D_k F_{ij}=D_i D_m F_{jk}+ {{R_{mi}}^l}_j F_{lk} +{{R_{mi}}^l}_k F_{jl}+ D_m D_j F_{ki} + D_m D_k F_{ij}
 \eeaa
  Contracting again with $D^i$ and commuting covariant derivatives, we obtain
  \beaa
 0=D^i D_i D_m F_{jk}+D^i{{R_{mi}}^l}_j F_{lk} +{{R_{mi}}^l}_j D^iF_{lk} +D^i {{R_{mi}}^l}_k F_{jl}+ {{R_{mi}}^l}_k D^i F_{jl}+ D^i D_m D_j F_{ki} + D^i D_m D_k F_{ij}
 \eeaa
 The last two terms can be reduced to 
 $$D^i D_m D_j F_{ki}=D_m D^i D_j F_{ki}+{{R^i}_{mj}}^l D_l F_{ki}+{{R^i}_{mk}}^l D_j F_{li}+{R_m}^l D_j F_{kl}$$ and commuting again derivatives in the first term,
 $$D_m(D^i D_j F_{ki})=D_m(D_j (D^i F_{ki}) + {{R^i}_{jk}}^l F_{li} + {R_j}^l F_{kl})$$ and since the divergence term cancels out, the dependence in $F$ is only up to one derivative. We obtain the desired expression.
  \end{proof}

 We will use the notation $\mathcal{M}(B^1, \dots, B^k)$ to denote a combination of contractions of tensors $B^1, \dots, B^k$, i.e. a tensor of the form 
 \bea
 \label{definitionM-general}
 \mathcal{M}(B^1, \dots, B^k)= \sum_{i, j} B^i \cdot B^j 
  \eea
where $\cdot$ denotes a contraction of tensors. We allow $B^i=B^j$, and in the notation $\mathcal{M}(B^1, \dots, B^k)$ we only write the different $B$s. 
 
 For instance, by combining \eqref{Einsteineq} and \eqref{def-stress-energy}, we have
 \beaa
 \operatorname{Ric}&=& 2g^{lk}F_{il}F_{jk}-\frac 1 2 g_{ij} (g^{lm}g^{kp}F_{lk}F_{mp})=\mathcal{M}( F) \\
 D\operatorname{Ric}&=&\mathcal{M}(F, DF) \\
DD\operatorname{Ric}&=&\mathcal{M}(F, DF, DDF)
 \eeaa
Using the above notation, and the fact that we can translate the dependence on $\operatorname{Ric}$, D$\operatorname{Ric}$, DD$\operatorname{Ric}$ into a dependence on $F, DF, DDF$, Proposition \ref{boring-relations} can be summarized as
 \bea
 \Box_g F&=& \mathcal{M}(F, \operatorname{Ric}, \operatorname{Riem}) = \mathcal{M}(F, \operatorname{Riem}) \label{waveFM}\\
 \Box_g \operatorname{Ric}&=& \mathcal{M}(F, DF, \operatorname{Ric}, \operatorname{Riem})=\mathcal{M}(F, DF, \operatorname{Riem})  \label{waveRicciM}\\
 \Box_g \operatorname{Riem}&=& \mathcal{M}(\operatorname{Ric}, DD\operatorname{Ric},  \operatorname{Riem})= \mathcal{M}(F, DF, DDF,  \operatorname{Riem}) \label{waveRiemannF} \\
 \Box_g DF&=& \mathcal{M}(F, DF, \operatorname{Ric}, D\operatorname{Ric}, \operatorname{Riem},  D\operatorname{Riem})=\mathcal{M}(F, DF, \operatorname{Riem},  D\operatorname{Riem})  \label{waveDFM}
 \eea
The above simplification will be used later.
 
\section{Unique continuation of the Killing vector field}\label{unique-continuation}

The aim of this section is to prove Theorem \ref{Theorem}. 

Let $(M, g)$ be an electrovacuum spacetime. 
Let $Z$ be a vector field defined on any open set $O$ of $M$. We recall that $Z$ is called a Killing vector field if the Lie derivative of the metric tensor $g$ of $M$ with respect to $Z$ vanishes.
 We assume in addition that the Killing vector field $Z$ preserves the electromagnetic tensor $F$ in the open domain $O$, i.e. 
  \bea
\label{Killingeq}
\mathcal{L}_Z g=0, \ \ \mathcal{L}_Z F=0 \ \ \text{in $O$}
\eea 

We also recall by standard arguments (see for example Lemma 3 of \cite{Nomizu}) that a Killing vector field defined on a connected open set is uniquely determined by its values and the value of its covariant derivative at any single point of the connected open set.\footnote{This fact is implied by the Killing equation being a second order ordinary differential equation for the vector field.}

\begin{defn}\label{definingfunction} A domain $O \subset M$ is said to be strongly null convex at a boundary point $p \in \partial O$ if it admits a defining smooth function\footnote{We observe that this definition does not depend on the defining function $h$, and it is automatically satisfied if the metric $g$ is Riemannian. It is satisfied also when the metric is Lorentzian and $\partial O$ is spacelike. On the other hand, it is never satisfied if $\partial O$ is null. This condition imposes interesting restrictions only when the hypersurface is timelike.
} $h:U \to \mathbb{R}$ with $U$ an open neighborhood of $p$ in $M$, $D h(p) \neq 0$ such that $O \cap U=\{ x \in U: h(x)<0\}$ and $$D^2 h(X,X)(p)<0$$ for any $X \neq 0 \in T_p(M)$ for which $g(X,X)=g(X,D h)=0$.
\end{defn}

\

We will now overview the proof of Theorem \ref{Theorem}.

We extend the vectorfield $Z$ along a geodesic, generated by a vector field denoted by $L$, using the Jacobi equation. This is done in Section \ref{extension-section}. Our aim is to prove that the extended vectorfield remains indeed a Killing field in a full neighborhood of the point $p$. 

To prove it, we show that the deformation tensor associated to $Z$, denoted as $\pi=\mathcal{L}_Z g$, vanishes identically for the new extended vector field. However, this is not enough to prove unique continuation: we need to show the simultaneous vanishing of the tensor $\pi$, of the Lie derivative of the Riemann tensor and the Lie derivative of the electromagnetic tensor. 

In order to do so, we obtain a transport equation for the tensor $\pi$ along the geodesic, and  we derive wave equations for the additional tensors, coupled with a certain number of transport equations for tensors who are all supposed to vanish in the case of a Killing vector field. This is done in Section \ref{waves-section}.

We close the system of transport and wave equations and, because of the null convexity condition, we are able to apply Carleman estimates to the system to obtain unique extension. This is done in Section \ref{proof-section}.

\subsection{Extension of the vector field}\label{extension-section}

Recall that the restriction of a Killing field to a geodesic is a Jacobi field. Inspired by this property, we extend the Killing vector field $Z$ along a geodesic using the Jacobi equation, as previously done in \cite{ExtensionRicciFlat}.

 Consider a geodesic which is transversal to $\partial O$ defined in a neighborhood $U$ of $p$ in $M$ and generated by a vectorfield $L$. In particular, $L$ is a smooth vector field defined in a neighborhood of $p$ such that $D_L L=0$ and $L(f)(p)=0$. We use the Jacobi equation\footnote{Recall that the Jacobi equation is a second order ODE, and therefore it admits a local solution.}  to extend $Z$ in $U \setminus O$, i.e. we solve in a neighborhood of $p$,
 \bea
\label{Jacobi}
D_L D_L Z=R(L,Z)L
\eea
with the obvious initial conditions. We still denote the extension $Z$. Therefore, after restricting the neighborhood $U$ of $p$ in $M$, we may assume that $Z$ and $L$ are smooth vector fields verifying 
\bea
D_L L&=&0, \qquad D_L D_L Z=R(L,Z)L \qquad \text{in $U$} \label{def-extension-eqs}\\
\mathcal{L}_Z g&=&0, \qquad \mathcal{L}_Z F=0 \qquad \text{in $O$} \label{identities-lie-derivatives}
\eea
We want to prove that the identities \eqref{identities-lie-derivatives} are satisfied in $U$ too. 

We summarize here the definition\footnote{We follow the notations in \cite{ExtensionRicciFlat}} of few tensors which are involved in the equations below. 
\begin{itemize} 
\item $B=\frac 1 2 (\pi + \omega)$, where $\omega$ is a $2$-form to be determined later,
\item $W= \mathcal{L}_Z \operatorname{Riem} -B \odot \operatorname{Riem}$, where $\operatorname{Riem}$ is the (0,4) Riemann tensor,  and $\odot$ is the Nomizu product of a $(0,2)$ tensor and a $(0,4)$ tensor\footnote{The Nomizu product of a $(0,2)$ tensor $B$ and a $(0,4)$ tensor $R$ is defined to be the $(0,4)$ tensor $(B\odot R)_{ijkl}={B_i}^m R_{mjkl}+{B_j}^m R_{imkl}+{B_k}^m R_{ijml}+{B_l}^m R_{ijkm}$.}, 
\item $E=\mathcal{L}_Z F- B \odot F$, where $\odot$ is the Nomizu product between two $(0,2)$ tensors\footnote{The Nomizu product of two $(0,2)$ tensor $B$ and $R$ is defined to be the $(0,2)$ tensor $(B\odot R)_{ij}={B_i}^m R_{mj}+{B_j}^m R_{im}$.},
\item $G=\mathcal{L}_Z (DF) - B \odot DF$ where $\odot$ is the Nomizu product of a $(0,2)$ tensor and a $(0,3)$ tensor\footnote{The Nomizu product of a $(0,2)$ tensor $B$ and a $(0,3)$ tensor $A$ is defined to be the $(0,3)$ tensor $(B\odot A)_{ijk}={B_i}^m A_{mjk}+{B_j}^m A_{imk}+{B_k}^m A_{ijm}$.},
\item $\dot{B}=D_L B$,
\item $P_{ijk}= D_i \pi_{jk}-D_k\pi_{ij}-D_j\omega_{ik}$
\end{itemize}
The above tensors all depend on the $2$-form $\omega$, which will be defined later, see equation \eqref{omega}.

Observe that the tensors $W$ and $E$ are very natural to consider. Indeed, in order to show that the extended vector field $Z$ is Killing, we want to prove that $\pi$ and $E$ are identically zero. Moreover, if $Z$ is Killing, then $\mathcal{L}_Z \operatorname{Riem}=0$, so $W$ vanishes too. The tensor $P$ appears in the computations of the wave equations for $W$ and $E$. 

The tensor $G$ seems redundant in the list of the above tensors. It is nevertheless necessary. Indeed, our goal is to apply Carleman estimates to a system of wave equations involving first covariant derivatives only. However, notice that the Riemann tensor verifies a wave equation that depends on the second covariant derivative of $F$, see equation (\ref{waveRiemannF}). For this reason, we introduce the tensor $G$, which is the Lie derivative of $DF$. The tensor $DF$ has the fundamental property that it verifies a wave equation depending on the first covariant derivative of $F$ only, see equation (\ref{waveDFM}). This allows us to write a system with one additional equation - the wave equation for $G$ - but depending only on the first covariant derivatives of the tensors which are involved.

In the case of Ricci flat manifolds treated in \cite{ExtensionRicciFlat}, the wave equation of the Riemann tensor does not depend on any of its covariant derivatives, therefore the argument works with the only use of the tensor $W$.

\subsection{Transport and wave equations}\label{waves-section}
We recall the following lemma.
\begin{lem}[Lemma 7.1.3 in \cite{stabilityMinkowski}]\label{Gamma} For any (0,k) tensor $V$ and a vectorfield $X$, we have 
\beaa
D_i \mathcal{L}_X V_{j_1 \dots j_k}- \mathcal{L}_X D_i V_{j_1 \dots j_k}=\sum_{n=1}^{k}\Gamma_{j_n i m}{(V_{(n)}^m)}_{j_1 \dots j_k}
\eeaa
 where ${(V_{(n)}^m)}_{j_1 \dots j_k}= V_{j_1 \dots \ \dots j_k}^{ \ \ \ m}$ is the tensor $V$ with the $n$-th index raised up using the metric, $\Gamma$ is defined as $\Gamma_{jim}=\frac 1 2 (D_j \pi_{im}+D_i \pi_{jm}-D_m \pi_{ij}),$ with $\pi$ is the deformation tensor of $X$. 
\end{lem}

We make use of the above Lemma to derive the wave equations for the tensors $E$, $G$ and $W$. 

 We first introduce the notation $\mathcal{M}_g(B^1, \dots, B^k)$ to denote a combination of contractions of tensors $B^1, \dots, B^k$ allowing coefficients depending on $g$, $\operatorname{Riem}$, $\operatorname{Ric}$, $F$ or their covariant derivatives, i.e. a tensor of the form 
 \bea
 \label{definitionM-general}
 \mathcal{M}_g(B^1, \dots, B^k)= \sum D^{\leq1}(g+\operatorname{Riem}+\operatorname{Ric}+F)^i \cdot B^j
  \eea
where $\cdot$ denotes a contraction of tensors. In the notation $\mathcal{M}_g(B^1, \dots, B^k)$ we only write the $B$s which are different from $g$, $\operatorname{Riem}$, $\operatorname{Ric}$, $F$ or their covariant derivatives. 

For example, by definition of $E$ we can write
\beaa
E&=&\mathcal{L}_Z F- B \odot F=\mathcal{M}(\mathcal{L}_Z F, B, F) = \mathcal{M}_g(\mathcal{L}_Z F, B)
\eeaa

\begin{prop}\label{waveequations} The following schematic wave equations hold:
\beaa
\Box_g E&=&\mathcal{M}_g( E, B, DB, DP, W) \\
\Box_g G&=&\mathcal{M}_g(E, B, DB, DP,  G, W, DW ) \\
\Box_g W&=&\mathcal{M}_g(E, B, DB, DP, G, DG, W)
\eeaa
\end{prop}
\proof We compute the wave equation of $E$. Applying Lemma \ref{Gamma}, we compute
\beaa
\Box_g \mathcal{L}_Z F_{ij}&=& D^k D_k \mathcal{L}_Z F_{ij}= D^k(\mathcal{L}_Z D_k F_{ij} +\Gamma_{ikm}{F^m}_j + \Gamma_{jkm}{F_i}^m)\\
&=& \mathcal{L}_Z (\Box_g F_{ij})+{{\Gamma_{k}}^k}_m D^mF_{ij}+{{\Gamma_{i}}^k}_m D_k{F^m}_{j}+{{\Gamma_{j}}^k}_m D_k{F_{i}}^m \\
&&+ D^k\Gamma_{ikm}{F^m}_j + \Gamma_{ikm}D^k{F^m}_j + D^k\Gamma_{jkm}{F_i}^m+\Gamma_{jkm}D^k{F_i}^m
\eeaa
Using \eqref{waveFM}, we can write 
\beaa
\mathcal{L}_Z (\Box_g F_{ij})&=& \mathcal{L}_Z (\mathcal{M}(F, \operatorname{Riem}))=\mathcal{M}(F, \operatorname{Riem}, \mathcal{L}_ZF, \mathcal{L}_Z\operatorname{Riem})=\mathcal{M}_g( \mathcal{L}_ZF, \mathcal{L}_Z\operatorname{Riem})=\mathcal{M}_g( E, B, W)
\eeaa
Observe that we can also write $\Gamma=\mathcal{M}_g( B, DB)$. We therefore simplify
\bea\label{first-part-wave-eq-E}
\Box_g \mathcal{L}_Z F_{ij}&=&  D^k\Gamma_{ikm}{F^m}_j  + D^k\Gamma_{jkm}{F_i}^m+\mathcal{M}_g( E, B, DB, W) 
\eea
The second part of the equation for $E$ is given by
\beaa
\Box (B \odot F)_{ij}&=& D^k D_k ({B_i}^m F_{mj} + {B_j}^m F_{im})=D^k (D_k{B_i}^m F_{mj}+{B_i}^m D_kF_{mj} +D_k{B_j}^m F_{im}+ {B_j}^m D_kF_{im})\\
&=& D^k D_k{B_i}^m F_{mj} +2D_k{B_i}^m D^kF_{mj} +{B_i}^m \Box_g F_{mj}\\
&&+D^k D_k{B_j}^m F_{im}+ 2D_k{B_j}^m D^kF_{im}+{B_j}^m \Box_gF_{im}
\eeaa 
Using \eqref{waveFM} again, we obtain
\bea\label{second-part-wave-eq-E}
\Box (B \odot F)_{ij}&=& D^kD_k{B_i}^m F_{mj} +D^kD_k{B_j}^m F_{im}+\mathcal{M}_g(B, DB)
\eea
Putting together \eqref{first-part-wave-eq-E} and \eqref{second-part-wave-eq-E}, we obtain
\beaa
\Box_g E_{ij}&=&\Box_g \mathcal{L}_Z F_{ij}-\Box_g (B \odot F)_{ij}\\
&=& D^k\Gamma_{ikm}{F^m}_j  + D^k\Gamma_{jkm}{F_i}^m-D^kD_k{B_i}^m F_{mj} -D^kD_k{B_j}^m F_{im}+\mathcal{M}_g( E, B, DB, W) \\
&=& D^k\left(\Gamma_{ikm}-D_kB_{im}\right){F^m}_j  +D^k\left(\Gamma_{jkm}-D_kB_{jm}\right) F_{i}^m+\mathcal{M}_g( E, B, DB, W) 
\eeaa
By definition of $P$, we obtain $\Gamma_{ikm}-D_kB_{im}=\frac 1 2 P_{imk}$, finally giving 
\beaa
\Box_g E_{ij}&=& \frac 1 2 D^kP_{imk}{F^m}_j  +D^kP_{jmk} F_{i}^m+\mathcal{M}_g( E, B, DB, W)=\mathcal{M}_g( E, B, DB, DP, W)
\eeaa
as desired.

We similarly compute the wave equations for $G$ and $W$. We show the main passages. Applying Lemma \ref{Gamma}, we compute
\beaa
\Box_g \mathcal{L}_Z (DF)_{i_1i_2i_3}&=&D^k(\mathcal{L}_Z D_k (DF)_{i_1i_2i_3} +\sum_{j=1}^3 \Gamma_{i_jkm}{(DF)_{(j)}^m}_{i_1 \dots i_3})\\
&=& \mathcal{L}_Z(\Box_g DF)+\sum_{j=1}^{3}{{\Gamma_{i_j}}^k}_m{(D_kDF_{(j)}^m)}_{i_j \dots i_3}\\
&&+ \sum_{j=1}^3 (D^k\Gamma_{i_jkm}{(DF)_{(j)}^m}_{i_1 \dots i_3}+\Gamma_{i_jkm}D^k{(DF)_{(j)}^m}_{i_1 \dots i_3})
\eeaa
Using \eqref{waveDFM}, we can write
\beaa
\mathcal{L}_Z(\Box_g DF)&=& \mathcal{L}_Z (\mathcal{M}(F, DF,  \operatorname{Riem}, D\operatorname{Riem}))=\mathcal{M}(F, DF,  \operatorname{Riem}, D\operatorname{Riem}, \mathcal{L}_ZF, \mathcal{L}_ZDF,  \mathcal{L}_Z\operatorname{Riem}, \mathcal{L}_ZD\operatorname{Riem})\\
&=&\mathcal{M}_g(\mathcal{L}_ZF, \mathcal{L}_ZDF, \mathcal{L}_Z\operatorname{Riem}, \mathcal{L}_ZD\operatorname{Riem})=\mathcal{M}_g(E, B, G, W, DW )
\eeaa
We therefore obtain
\beaa
\Box_g \mathcal{L}_Z (DF)_{i_1i_2i_3}&=&\sum_{j=1}^3 (D^k\Gamma_{i_jkm}{(DF)_{(j)}^m}_{i_1 \dots i_3})+ \mathcal{M}_g(E, B, DB, G, W, DW )
\eeaa
The second part of the equation is given, as above, by
\beaa
\Box (B \odot DF)_{i_1i_2i_3}=\sum_{j=1}^3 D^k D_k B_{i_j m}{(DF)_{(j)}^m}_{i_1 \dots i_3} +\mathcal{M}(B, DB)
\eeaa
and the definition of $P$ allows to conclude as before. The wave equation for $W$ goes in the same way. In particular we obtain, using \eqref{waveRiemannF},  
\beaa
\mathcal{L}_Z(\Box_g \operatorname{Riem})&=& \mathcal{L}_Z (\mathcal{M}(F, DF, DDF \operatorname{Riem}))=\mathcal{M}(F, DF, DDF,  \operatorname{Riem}, \mathcal{L}_ZF, \mathcal{L}_ZDF, \mathcal{L}_ZDDF,  \mathcal{L}_Z\operatorname{Riem})\\
&=&\mathcal{M}_g(\mathcal{L}_ZF, \mathcal{L}_ZDF, \mathcal{L}_ZDDF,  \mathcal{L}_Z\operatorname{Riem})=\mathcal{M}_g(E, B, G, DG, W)
\eeaa
and using the tensor $P$ as before, we obtain the desired expression. 
\endproof
 
The transport equations for the tensors $B$, $\dot{B}$ and $P$ are consequences of the extension \eqref{def-extension-eqs}. In particular, they do not use Einstein equation. They are therefore identical to the ones in the case of Ricci flat manifolds, as obtained in \cite{ExtensionRicciFlat}. 
\begin{lem}[Lemma 2.6 and Lemma 2.7 in \cite{ExtensionRicciFlat}]\label{transportequations} Given the vectorfield $Z$, extended to $M$ by (\ref{Jacobi}), we have 
\beaa
L^i \pi_{ij}=0 \qquad \text{in $M$}
\eeaa 
Moreover, if we define $\omega$ in $M$ as the solution of the transport equation 
\bea
\label{omega}
D_L \omega_{ij}= \pi_{ik}D_j L^k - \pi_{jk} D_i L^k
\eea
with $\omega=0$ in $O$, then 
\beaa
L^k P_{ijk}=0, \qquad L^i \omega_{ij}=0 \qquad \text{in $M$}.
\eeaa

In $M$ we have
\beaa
 D_L B_{ij}&=& \dot{B}_{ij} \\
D_L \dot{B}_{ij}&=& L^k L^m (\mathcal{L}_Z R)_{kijm} - 2 \dot{B}_{kj}D_iL^k - {\pi_j}^k L^m L^p R_{mikp} \\ 
 D_L P_{ijk}&=& 2 L^m W_{ijkm} + 2 L^m {B_k}^p R_{ijpm} - D_k L^m P_{ijm}
 \eeaa
\end{lem}

We can summarize Lemma \ref{transportequations} as 
\beaa
D_L B&=& \mathcal{M}_g(\dot{B}) \\
D_L \dot{B}&=& \mathcal{M}_g(B, \dot{B}, W) \\
D_L P&=& \mathcal{M}_g(B, P, W)
\eeaa

As a consequence of Proposition \ref{waveequations} and Lemma \ref{transportequations}, we have obtained the following system of equations in $M$:
\bea 
\label{system2}
\left\{
                \begin{array}{ll}
                  \square_g S^i=\mathcal{M}(B^1, \dots, B^k, S^1, DS^1, \dots, S^d, DS^d), \   \  i= 1, \dots, d \\
                  D_L B^j= \mathcal{M}(B^1, \dots, B^k,   S^1, DS^1, \dots, S^d, DS^d), \  \  j=1, \dots, k 
                \end{array}
                \right.
\eea
where $S^i=E, G, W$ and $B^j=B, DB, \dot{B}, D\dot{B}, P, DP$.

Our goal is to prove that solutions $S^1, \dots, S^d, B^1, \dots, B^k$ to the above system which vanish on one side of a hypersurface verifying the strong null convexity condition have to vanish in a full neighborhood of the hypersurface.

 Notice that the system is completely analogous to the one obtained in \cite{ExtensionRicciFlat}, except that in here, due to the coupling with the electromagnetic tensor, we have more than one wave equation.

\subsection{Carleman estimates and proof of Theorem \ref{Theorem}}\label{proof-section}

We are now ready to prove Theorem \ref{Theorem}. 

Upon introducing a system of coordinates around the point $p$ in which the metric $g$ takes the form $g_{ij}(p)=\operatorname{diag}(-1, \dots, -1, 1, \dots, 1)$,  we obtain a set of functions $G_i: B_{\delta_0} \to \mathbb{C}$, $i=1, \dots I$, for the tensors $S^1, \dots, S^d$, and a set of functions $H_j: B_{\delta_0} \to \mathbb{C}$, $j=1, \dots J$, for the tensors $B^1, \dots, B^k$, that in view of equation (\ref{system2}) verify
\bea
\label{systemcoordinates}
\begin{split}
|\square_g G_i | \leq M \sum_{l=1}^I (|G_l|+|\partial^1 G_l|) + M \sum_{m=1}^J |H_m| \\
|L(H_j)| \leq M \sum_{l=1}^I(|G_l|+|\partial^1 G_l|)+ M \sum_{m=1}^J |H_m|
\end{split}
\eea
for any $i=1, \dots I$, $j=1, \dots J$, where $M \geq 1$ is a constant. Therefore, Lemma 2.10 of \cite{ExtensionRicciFlat} whose proof is based on Carleman estimates for the equations \eqref{system2}, applies identically in this case to imply Theorem \ref{Theorem}. We summarize it in the following lemma.
\begin{lem}\label{LemmaCarleman} Assume that $\delta_0 > 0$ and $G_i, H_j : B_{\delta_0} \to \mathbb{C}$ are smooth functions, $i=1, \dots I$, $j=1, \dots J$ such that they satisfy equations (\ref{systemcoordinates}) for a costant $M \geq 1$. Assume that $G_i=0$ and $H_j=0$ in $B_{\delta_0}(p) \  \cap \ O$, $i=1, \dots I$, $j=1, \dots J$. Assume also that $h$ is strongly null convex at $p$, in the sense of Definition \ref{definingfunction}, and $L(h)(p) \neq 0$. Then $G_i=0$ and $H_j=0$ in $B_{\delta_1}(p)$, $i=1, \dots I$, $j=1, \dots J$, for some constant $\delta_1 \in (0, \delta_0)$ sufficiently small.
\end{lem}
The above Lemma implies that all the tensors defined in Section  \ref{extension-section} for the extended vectorfield $Z$ vanish identically. Therefore $Z$ is indeed a Killing vectorfield which preserves the electromagnetic tensor, i.e. $$\mathcal{L}_Z g=0, \ \ \ \mathcal{L}_Z F=0$$ in a full neighborhood of $p$. This completes the proof of the theorem. 

\section{Counterexample to the extendibility}\label{counterexample-section}
Our second theorem provides a counterexample to extendibility, in the setting of the charged black hole rigidity problem. 

Let $(\mathcal{K}_{m,a,Q}, g)$ denote the Kerr-Newman spacetime of mass $m$, angular momentum $a$ with $0<a<m$, and charge $Q$ with $0<Q<m$. Let $\mathcal{D}$ denote the domain of outer communication, and $\mathcal{H}^+$ the event horizon of the black hole. Let $T=\partial_t$ denote the stationary Killing vector field of Kerr-Newman, and let $Z=\partial_{\phi}$ denotes its axially symmetric Killing vector field.

We state here the precise version of the theorem containing the counterexample to extendibility.

\begin{thm}
\label{Counterexample}
Let $U_0\subset \mathcal{K}_{m,a,Q}$, with $0<a, |Q|<m$, be an open set such that $U_0 \cap \mathcal{H}^+ \cap \overline{\mathcal{D}}\neq \emptyset$. Then there is an open set $U \subset U_0$ diffeomorphic to the open ball $B_1 \subset \mathbb{R}^4$, $U \cap \mathcal{H}^+ \neq \emptyset$ and a smooth Lorentz metric $\tilde{g}$ in $U$ with the following properties:
\begin{itemize}
\item $(U, \tilde{g})$ is an electrovacuum spacetime,
\item $\mathcal{L}_{T} \tilde{g}=0$ in $U$, 
\item $g=\tilde{g}$ and $F=\tilde{F}$ in $U \setminus \mathcal{D}$,
\item the vector field $Z$ does not extend to a Killing vector field for $\tilde{g}$ in $U$, commuting with $T$.
\end{itemize}
\end{thm}

The above counterexample shows the necessity of the strongly null convex condition for the extension of a Killing field as in Theorem \ref{Theorem}. The proof follows the same procedure of the counterexample constructed in Theorem 1.3. in \cite{ExtensionRicciFlat} in the case of Kerr spacetime.

\

We will now overview the proof of Theorem \ref{Counterexample}.

We first recall the main properties and important quantities in Kerr-Newman spacetime. This is done in Section \ref{calculations-kerr-newman}.

Fix a point $p \in U_0 \cap \mathcal{H}^+ \cap \overline{\mathcal{D}}$ outside the bifurcation sphere $S_0=\mathcal{H}^+ \cap \mathcal{H}^-$ and the axis of symmetry $A=\{ p \in \overline{\mathcal{D}} : Z(p)=0 \}$. Consider the Kerr-Newman metric $g$ and the induced metric $$h_{ij}= X g_{ij} - T_i T_j,$$ where $X=g(T,T)$, on a hypersurface $\Sigma$ passing through the point $p$ and transversal to $T$. 
\begin{rem}
The metric $h$ is nondegenerate as long as $X=g(T,T) >0$ in $\Sigma$, i.e. as long as the vector field $T$ is spacelike. Therefore we have to perform this construction inside the ergoregion. 
\end{rem}
 The construction of the counterexample makes use of the reduced equations in the induced metric in $\Sigma$, and solves the characteristic problem for them. The Einstein-Maxwell equation, together with the stationarity condition $\mathcal{L}_T g=0$, are equivalent to the following system of equations in $\Sigma$ (derived in Appendix \ref{appendix})
\bea
\begin{split}
\label{reducedsystem}
{\operatorname{Ric}^{(h)}}_{ij}&=\operatorname{Ric}_{ij} + \frac 1 2 X^{-1} \nabla_i X \nabla_j X - X^{-1} {B_j}^m B_{im}, \\
\square_h \sigma&= - X^{-1}(D^i \sigma D_i \sigma + \frac 1 2 D^i X D_i \sigma) + 3 E^2, \\
\square_h \psi &= -B^{ij} \mathcal{F}_{ij} - X^{-1} D^i X D_i \psi
\end{split}
\eea
where $\sigma$ is the Ernst potential associated to $T$ and $\psi$ is the electromagnetic potential associated to $T$. The remaining quantities involved in the system above are defined in the Appendix.

We modify the metric $h$ and the functions $X$ and $Y$\footnote{The functions $X$ and $Y$ are defined to be the negative real and imaginary part of the Ernst potential $\sigma$.} in a neighborhood of the point $p$ in such a way that the identities (\ref{reducedsystem}) are still satisfied. The existence of a large family of smooth data $(\tilde{h}, \tilde{\sigma}, \tilde{\phi})$ satisfying (\ref{reducedsystem}) and agreeing with the Kerr-Newman data in $\Sigma \setminus \mathcal{D}$ follows by solving a characteristic initial value problem (see for example \cite{Rendall}).

We construct the spacetime metric $\tilde{g}$ associated to the induced metric $\tilde{h}$ as 
\bea
\label{gfromreduced}
\tilde{g}_{ij}=\tilde{X}^{-1} \tilde{h}_{ij} + \tilde{X} \tilde{A}_i \tilde{A}_j, \ \ \tilde{g}_{i4}= \tilde{X} \tilde{A}_i, \ \ \tilde{g}_{44}=\tilde{X}
\eea
for $T= \partial_4$, where $\tilde{A}$ is a suitable $1$-form, which verifies the following compatibility equation, as derived in the Appendix:
\bea
\label{equationA}
\nabla_i \tilde{A}_j - \nabla_j \tilde{A}_i = \tilde{X}^{-3/2}({\epsilon^{(h)}}_{ijk} \nabla^k \tilde{Y})
\eea
 By construction, such a metric verifies the Einstein-Maxwell equations and $\mathcal{L}_T \tilde{g}=0$ in a suitable open set $U$.
Finally, we show that we have enough flexibility to choose initial conditions for $\tilde{X}, \tilde{Y}$ such that the vector field $Z$ cannot be extended as a Killing vector field for $\tilde{g}$ in the open set $U$.  This is done in Section \ref{construction-section}.

\subsection{Explicit calculations in Kerr-Newman spacetime}\label{calculations-kerr-newman}

The Kerr-Newman spacetime $\mathcal{K}_{m,a,Q}$ in Boyer-Lindquist coordinates is given by 
\beaa
g&=& -\left(1- \frac{2mr - Q^2}{\rho^2}\right)dt^2 - \frac{2a(2mr-Q^2)}{\rho^2} \sin^2 \theta d\phi dt+ \sin^2\theta\left(r^2 + a^2 +\frac{a^2(2mr-Q^2)}{\rho^2}\sin^2\theta\right)d\phi^2 + \frac{\rho^2}{\Delta} dr^2 + \rho^2 d\theta^2
\eeaa
where 
\beaa
\rho^2=r^2 + a^2 \cos^2\theta, \qquad \Delta=r^2 + a^2 -2mr + Q^2.
\eeaa
 Making the change of variables 
 \beaa
 du_{-}=dt-\frac{r^2+a^2}{\Delta} dr, \qquad  d\phi_{-}=d\phi - \frac{a}{\Delta} dr
 \eeaa
  then the spacetime in the new coordinates $(\theta, r, \phi_{-}, u_{-})$ becomes 
  \bea\label{metric-g-new-coords}
  g= \rho^2 d\theta^2 - 2 du_{-} dr + 2a\sin^2 \theta d\phi_{-} dr - \frac{ 2a (2mr-Q^2)\sin^2\theta}{\rho^2} d\phi_{-} du_{-} +\frac{ \Sigma^2 \sin^2 \theta}{\rho^2} d\phi_{-}^2 + \frac{2mr-Q^2-\rho^2}{\rho^2} du_{-}^2
  \eea
   where 
   \beaa
   \Sigma^2= (r^2+a^2)\rho^2 + a^2(2mr-Q^2)\sin^2\theta=(r^2+a^2)^2-a^2 \sin^2\theta \Delta
   \eeaa
    The vectorfield $T=\partial_t$ becomes $T=\partial_{u_{-}}$ upon change of coordinates. The metric $g$ and the vector field $T$ are smooth in the region $R=\{ (\theta, r, \phi_{-}, u_{-})\in (0,\pi) \times (0, \infty) \times (-\pi, \pi) \times \mathbb{R} : 2mr -Q^2 - \rho^2 >0\}.$

Let 
\beaa
X= g(T,T)= \frac{2mr-Q^2-\rho^2}{\rho^2}
\eeaa
 and let $h_{ij}= X g_{ij} - T_i T_j$ be the induced metric on $\Sigma=\{ (\theta, r, \phi_{-}, u_{-})\in R: u_{-}=0 \}$. Let $\partial_1 = \partial_{\theta}$, $\partial_2= \partial_r$, $\partial_3=\partial_{\phi_{-}}$, then the components of the metric $h$ along the surface $\Sigma$ are 
 \beaa
 h_{11}=2mr-Q^2-\rho^2, \qquad h_{12}=h_{13}=0, \qquad h_{22}=-1, \qquad h_{23}= - a \sin^2\theta, \qquad h_{33}=-\Delta \sin^2\theta
 \eeaa
 The Ricci curvature of the induced metric $h$ has components 
 \beaa
 {\operatorname{Ric}^{(h)}}_{11}&=& \frac{2m^2 a^2 \sin^2 \theta}{(2mr-Q^2-\rho^2)^2}, \qquad {\operatorname{Ric}^{(h)}}_{22}= \frac{2m^2}{(2mr-Q^2-\rho^2)^2}\\
&& {\operatorname{Ric}^{(h)}}_{12}={\operatorname{Ric}^{(h)}}_{13}={\operatorname{Ric}^{(h)}}_{23}={\operatorname{Ric}^{(h)}}_{33}=0.
 \eeaa
 The  spacetime metric $g$ in the coordinates $(\theta, r, \phi_{-}, u_{-})$ in \eqref{metric-g-new-coords} has the form 
 \beaa
 g_{ij}&=& X^{-1}h_{ij} + X A_i A_j, \qquad  g_{i4}=X A_i,\qquad \text{for $i, j=1,2,3$}  \\
  g_{44}&=&X
 \eeaa
  where 
  \beaa
  A_1&=&0, \qquad A_2= - \frac{\rho^2}{2mr-Q^2 - \rho^2}, \qquad A_3= - \frac{a \sin^2 \theta (2mr - Q^2)}{2mr-Q^2 -\rho^2}.
  \eeaa
 The electromagnetic potential of Kerr-Newman is given by
 \beaa 
 \psi= - \frac{Qr}{\rho^2} + i \frac{Q r a \sin^2 \theta}{\rho^2}
 \eeaa 
 The Ernst potential is given by
 \beaa 
 \sigma= 1- \frac{2mr-Q^2}{r(r+ia\cos\theta)}
 \eeaa
  and therefore 
  \beaa
  Y=-Im(\sigma)=-\frac{a \cos\theta (2mr-Q^2)}{r \rho^2}.
  \eeaa

A direct computation shows that the quantities $(h, \sigma, \psi, A)$ verify the system of equations (\ref{reducedsystem}) together with (\ref{equationA}).

\subsection{Construction of the counterexample}\label{construction-section}

As outlined above, in order to prove Theorem \ref{Counterexample}, we want to construct a family of smooth data $(\tilde{h}, \tilde{\sigma}, \tilde{\psi})$ and 1-forms $\tilde{A}$ which verify equations \eqref{reducedsystem} and \eqref{equationA} in a neighborhood in $\Sigma$ of $p$. 

We want to make advantage of the well-posedness of the system \eqref{reducedsystem} in the setting of the characteristic initial value problem (see for example \cite{Rendall}). In order to do so, we will construct the null hypersurface which is transversal to the horizon $\mathcal{H}^+$. We make use of the same construction used in \cite{ExtensionRicciFlat}.

 Let $N_0= \{ x \in \Sigma : r(x)= r_{+}= m + \sqrt{m^2 - a^2 - Q^2} \}$, where  $r_{+}$ is the largest root of $\Delta(r)=0$, and corresponds to the event horizon in Kerr-Newman spacetime. This is a 2-dimensional null hypersurface in $\Sigma$, with $\partial_1$ and $\partial_3$ tangent vector fields.
 
  Along $N_0$ we define the smooth transversal null vector field $L=(2a^2 \sin^2 \theta - \Delta)^{-1} (2 a \partial_2 - (\sin\theta)^{-2} \partial_3)$. Let $P=\{ x \in N_0: \phi_{-}(x)=0 \}$ and $p=\{ x\in P : \theta(x)=\theta_0 \in (0, \pi) \}$ for a certain $\theta_0$. Then, $P$ is a smooth curve in $N_0$ and $p \in P$ is a point on it. We extend the vector field $L$ to a small neighborhood $D$ of $p$ in $\Sigma$, by solving the geodesic equation $D_L L=0$ in $D$. Then we construct the null hypersurface $N_1$ in $D$ as the congruence of geodesic curves tangent to $L$ and passing through the curve $P$. Define $D_{-}= \{ x\in D: \Delta(x) <0 \}$ and $D_{+}= \{ x\in D: \Delta(x) >0 \}$.

The following proposition is a conseguence of the main theorem in \cite{Rendall}.
\begin{prop}
\label{construction1}
Assume $\tilde{\sigma}, \tilde{\psi} : N_1 \to \mathbb{C}$ are smooth functions satisfying $\tilde{\sigma}=\sigma$ and  $\tilde{\psi}=\psi $ in $N_1 \cap D_{-}$. Then there is a small neighborhood $D'$ of $p$ in $\Sigma$, a smooth metric $\tilde{h}$ in $J^{+}(N_1) \cap D'$, and smooth extension $\tilde{\sigma}, \tilde{\psi}:J^{+}(N_1) \cap D' \to \mathbb{C} $ such that in $J^{+}(N_1) \cap D'$,
\bea
\label{equationsneeded}
\begin{split}
{\operatorname{Ric}^{(\tilde{h})}}_{ij}&=\operatorname{Ric}_{ij} + \frac 1 2 \tilde{X}^{-1} \nabla_i \tilde{X} \nabla_j \tilde{X} - \tilde{X}^{-1} {\tilde{B}_j}^m \tilde{B}_{im}, \\
\square_{\tilde{h}} \tilde{\sigma}&= - \tilde{X}^{-1}(D^i \tilde{\sigma} D_i \tilde{\sigma} + \frac 1 2 D^i \tilde{X} D_i \tilde{\sigma}) + 3 \tilde{E}^2, \\
\square_h \tilde{\psi} &= -\tilde{B}^{ij} \tilde{\mathcal{F}}_{ij} - \tilde{X}^{-1} D^i \tilde{X} D_i \tilde{\psi}
\end{split}
\eea
where $\operatorname{Ric}_{ij}$ is written in terms of $\tilde{\psi}$ through the Einstein-Maxwell equations. 

 In addition, 
 \beaa
 \tilde{\sigma}=\sigma, \qquad  \tilde{\psi}=\psi, \qquad \tilde{h}=h \qquad \text{in $J^{+}(N_1) \cap D' \cap D_{-}$}
 \eeaa
\end{prop}
To construct the desired spacetime metric $\tilde{g}$ from the reduced one $\tilde{h}$ using the formula (\ref{gfromreduced}), we need to extend the 1-form $A$. This can be done by solving \eqref{equationA} for $\tilde{A}$, with $\tilde{A}=A$ in $D \cap D_{-}$ (see Proposition 3.4. in \cite{ExtensionRicciFlat}). We finally extend the functions $\tilde{h}_{ij}, \tilde{\sigma}, \tilde{\psi}, \tilde{A}_i$, originally defined in $D$, to $D \times I$ by the conditions $\partial_4(\tilde{\sigma})=\partial_4(\tilde{\psi})=\partial_4(\tilde{A_i})=\partial_4(\tilde{h}_{ij})=0.$

By construction, the metric $\tilde{g}$ agrees with the Kerr-Newman metric $g$ in $(D \cap D_{-}) \times I$ and satisfies the Einstein-Maxwell equations. In addition, it verifies
\beaa
\mathcal{L}_{\partial_4} \tilde{g}=0, \qquad  \mathcal{L}_{\partial_4} \tilde{F}=0  \qquad \text{in $D \times \mathbb{R}$}
\eeaa 
 Moreover if $Z= Z^4 \partial_4 + Z^i \partial_i$ is a Killing vector field for $\tilde{g}$  in $D \times I$ and if $[Z, \partial_4]=0$, then $Z'=Z^i \partial_i$ is a Killing vector field for $\tilde{h}$ in $D$ satisfying $Z'(\tilde{X})=Z'(\tilde{Y})=0$.

To complete the proof of the Theorem \ref{Counterexample}, it remains to show that we can arrange our construction in such a way that the vector field $Z$ cannot be extended as a Killing vector field for the modified metric $\tilde{g}$. In particular, it suffices to prove that we can arrange the construction such that the vector field $\partial_3$ (the one corresponding to the axial symmetric vector field $\phi$) cannot be extended to a vectorfield in $Z'$ in $D$ such that
\bea
\label{negation}
\mathcal{L}_{Z'} \tilde{h}=0, \ \ Z'(\tilde{X})=Z'(\tilde{Y})=0 \qquad \text{in $D$}
\eea
 As in \cite{ExtensionRicciFlat}, we assume that (\ref{negation}) holds and show that there is a choice of $\tilde{\sigma}$ and $\tilde{\psi}$ along $N_1$ such that (\ref{equationsneeded}) is violated. The proof follows the same pattern as in \cite{ExtensionRicciFlat}. We recall here the main steps.

Assuming that (\ref{negation}) holds, we define the geodesic vector field $\tilde{L}$ in $D$ as the geodesic vector field that coincides with $L$ on $((J^{+}(N_1)\cap D_{-})\cup N_1)\cap D'$, and choose a frame $e_2= \tilde{L}$, $e_3=Z'$ and $e_1$ an additional vector field in $D$ such that $\tilde{h}(e_1, e_2)=\tilde{h}(e_1, e_3)=0$, $\tilde{h}(e_1, e_1)=1$. We define the Ricci coefficients and the curvature components of the metric $\tilde{h}$ with respect to this frame. Indeed, we have 
\bea
\label{randomformulas}
\begin{split}
\tilde{h}(e_1, e_1)-1=\tilde{h}(e_1,e_2)=\tilde{h}(e_1,e_3)=\tilde{h}(e_2,e_2)=\tilde{h}(e_2,e_3)+1=0 \\
\Gamma_{i22}=0, \ \ \Gamma_{i3j}+\Gamma_{j3i}=0, \ \ \Gamma_{i3j}=\Gamma_{ij3} \\
{R^{\tilde{h}}}_{i323}=e_2(\Gamma_{i33})-\Gamma_{233}\Gamma_{i32}-\Gamma_{132}\Gamma_{i13}+\Gamma_{332}\Gamma_{i23} \\
{R^{\tilde{h}}}_{1223}=e_2(\Gamma_{123}), \ \ {R^{\tilde{h}}}_{2121}=e_2(\Gamma_{211})+\Gamma_{211}\Gamma_{121}
\end{split}
\eea
for any $i,j \in \{ 1,2, 3\}$. We can now obtain our desired contradiction by constructing a pair of smooth functions $\tilde{X}, \tilde{Y}$ that determine $\tilde{\sigma}$ along $N_1$ such that not all those identities can be simultaneously verified along $N_1$. Notice that in this electrovacuum case, we can simply leave unchanged the electromagnetic potential $\tilde{\psi}$, because the modification of $\tilde{\sigma}$ will already give the desired contradiction. We fix a smooth system of coordinates $y=(y^1, y^2, y^3)$ in a neighborhood of the point $p \in \Sigma$ such that $N_1=\{ q: y^3(q)=0\}$, $ N_0=\{q: y^2(q)=0\},$ and  $L=\tilde{L}=\partial_{y^2}$ along $N_1$. More precisely, we fix the $L$ as in the unperturbed Kerr-Newman in a neighborhood of $p$ and define first $y^2$ such that $y^2$ vanishes on $N_0$ and $L(y^2)=1$. Then we complete the coordinate system on $N_0$ and extend it by solving $L(y^1)=L(y^3)=0$. 

Assume $f: \mathbb{R}^3 \to [0,1]$ is a smooth function equal to $1$ in the unit ball and vanishing outside the ball of radius $2$. As in \cite{ExtensionRicciFlat}, we impose the following ansatz to $\tilde{X}, \tilde{Y}, \tilde{\psi}$: 
\bea
\label{ansatz}
\tilde{X}(q)=X(q), \qquad \tilde{Y}(q)= Y(q)+ \epsilon f\left(\frac{y(q) - y(p')}{\epsilon}\right), \qquad \tilde{\psi}(q)=\psi(q)
\eea
 for $q \in N_1$, where $p'$ is a fixed point in $N_1 \cap D_+$ sufficiently close to $p$, and $(X,Y, \psi)$ are the quantities in Kerr-Newman. We show that such a choice leads to a contradiction, for $\epsilon$ sufficiently small.

Let $e_i=K^j_i \partial_{y^i}$. Using the last identity in (\ref{randomformulas}) and the first identity in \eqref{equationsneeded} along $N_1$, we derive that
\bea
\label{firstrandom}
\partial_{y^2}(\Gamma_{211})-(\Gamma_{211})^2= \mathcal{M}(\tilde{\psi})+ \frac{1}{2\tilde{X}^2}[\partial_2(\tilde{X})^2 + \partial_2(\tilde{Y})^2] \qquad \text{along $N_1$}
\eea
where $\mathcal{M}(\tilde{\psi})$ denotes the dependence of the Ricci tensor on the electromagnetic potential given by the Einstein-Maxwell equations. In addition, since $[e_2, e_1]=\partial_{y^2}(K^1_1)\partial_{y^1}+\partial_{y^2}(K^2_1)\partial_{y^2}$ along $N_1$, it follows that $\partial_{y^2}(K^1_1)=K^1_1 \Gamma_{211}$ along $N_1$. Using the first identity in the last line in (\ref{randomformulas}) and the Ricci equation in \eqref{equationsneeded} similarly, together with the ansatz (\ref{ansatz}) it follows, by standard ODE theory, that $$|G| + |\partial_{y^2}(G)| \lesssim 1$$ for any $G \in \{ \Gamma_{211}, K^1_1, 1/K^1_1, \Gamma_{123}, K^2_1 \}$ along $N_1$, uniformly for all $p'\in N_1$ sufficiently close to $p$ and $\epsilon$ sufficiently small. 

Using the Ricci identity in \eqref{equationsneeded}, the identities $e_3(\tilde{X})=e_4(\tilde{X})=0$, and the bounds above, it follows that 
\beaa 
\sum_{i,j \in \{1,2,3\} } |{R^{\tilde{h}}}_{ij}| \lesssim 1
\eeaa
 along $N_1$. Using the identities in the first, second and third line of (\ref{randomformulas}), it follows that $$|\tilde{h}_{33}| + |\partial_{y^2}(\tilde{h}_{33})| + |\partial_{y^2}(\partial_{y^2}(\tilde{h}_{33}))| \lesssim 1$$ along $N_1$, uniformly for all $p' \in N_1$ sufficiently close to $p$ and $\epsilon$ sufficiently small. 

We can now derive a contradiction by examining the imaginary part of second equation in \eqref{equationsneeded} (see equation \eqref{waveeqY}):
\beaa
\tilde{h}^{ij} \tilde{\nabla}_i\tilde{\nabla}_j(\tilde{Y})= 2 \tilde{X}^{-1} \tilde{h}^{ij} e_i(\tilde{X}) e_j(\tilde{Y}).
\eeaa
 Using the previous bounds, it follows that $$|e_1(e_1(\tilde{Y}))-\tilde{h}_{33} e_2(e_2(\tilde{Y}))| \lesssim 1$$ along $N_1$, uniformly for all $p' \in N_1$ sufficiently close to $p$ and $\epsilon$ sufficiently small. This cannot happen as can be seen by letting first $\epsilon \to 0$ and then $p' \to p$, taking into account that $h_{33}=K^2_1=0$ along $N_0 \cap N_1$.

\appendix
\section{Reduction by Killing vector fields in electrovacuum}\label{appendix}
We derive the main equations used in the proof of Theorem \ref{Counterexample}, in full generality for a Killing vector field $K$.  In Section \ref{counterexample-section}, the equations below are applied to the stationary Killing vector field $T$.

Let $(M, g)$ be an electrovacuum spacetime, and $K$ be a Killing vector field, i.e. $D_i K_j + D_j K_i=0$. We also assume that $K$ preserves the electromagnetic tensor, i.e. $\mathcal{L}_K F=0$. 

We define the two form $B_{ij}=D_iK_j$ and $X=g(K,K)$. We recall that the Killing equation implies $D_i D_j K_l={R^m}_{ijl} K_m$. 
In view of the first Bianchi identity for $\operatorname{Riem}$, we infer that 
\bea 
D_i B_{jk} + D_j B_{ki}+ D_k B_{ij}=0 \label{BianchiA}\\
D^i B_{ji} = D^i D_j K_i= {R^{mi}}_{ji} K_m= {R^m}_j K_m \label{divergenceA}
\eea
Define the complex valued $2$-form $\mathcal{B}_{ij}= B_{ij} + i *B_{ij}$, where $*B$ is the Hodge dual of $B$. As a consequence of \eqref{BianchiA} and \eqref{divergenceA}, we obtain
\bea
 D_i \mathcal{B}_{jk} + D_j \mathcal{B}_{ki}+ D_k \mathcal{B}_{ij}=0, \\
D^i \mathcal{B}_{ji}= D^i B_{ji} + i D^i *B_{ji}= {R^i}_j K_i \label{divergence-1}
\eea
since $D^i *B_{ji}= \frac 1 2 \epsilon_{jikl} D^i B^{kl}= D_i B_{jk} + D_j B_{ki}+ D_k B_{ij}=0$.

We define the Ernst 1-form associated to the Killing vector field $K$ as 
\beaa
\sigma_i= 2 K^j \mathcal{B}_{ji}.
\eeaa
Observe that $D_i \sigma_j - D_j \sigma_i=0$. Indeed, 
\bea\label{dsigma=0}
2^{-1}(D_i \sigma_j - D_j \sigma_i)=K^m(D_j \mathcal{B}_{mi} -D_i \mathcal{B}_{mj}) + D_j K^m \mathcal{B}_{mi} - D_i K^m \mathcal{B}_{mj}= -\mathcal{L}_K \mathcal{B}_{ij}=0
\eea
We also have the following formula for the divergence of $\sigma$, using \eqref{divergence-1}:
\bea\label{divergence-sigma}
2^{-1} D^i \sigma_i = K^m D^i \mathcal{B}_{mi} + D^i K^m \mathcal{B}_{mi}= -2^{-1} \mathcal{B}^2 + R_{ij} K^i K^j
\eea

We introduce the following standard decomposition of the tensor $\mathcal{B}$. Suppose we have $$\textit{i}_K (B)_i=K^j B_{ji}, \ \ \textit{i}_K (*B)_i=K^j *B_{ji}, \ \ \textit{i}_K (\mathcal{B})_i=K^j \mathcal{B}_{ji}.$$ Then we can decompose $B$ and $\mathcal{B}$ as 
\bea
\label{decompositionA}
g(K,K) B_{ij}= K_i \textit{i}_K (B)_j - K_j \textit{i}_K (B)_i + \epsilon_{ijkl} K^k \textit{i}_K (*B)^l \\ 
g(K,K) \mathcal{B}_{ij}= K_i \textit{i}_K (\mathcal{B})_j - K_j \textit{i}_K (\mathcal{B})_i - i \epsilon_{ijkl} K^k \textit{i}_K (\mathcal{B})^l.
\eea
which can be written in terms of $\sigma$ as
\bea
\label{decompositionB}
2X B_{ij}= - K_i \sigma_j + K_j \sigma_i - \epsilon_{ijkl} K^k D^l Y \\
2 X \mathcal{B}_{ij}=-K_i \textit{i}_K \sigma_j + K_j \sigma_i + i \epsilon_{ijkl} K^k \sigma^l
\eea
In particular, 
\bea
\label{normsigma}
g(K,K) \mathcal{B}^2= 4 \textit{i}_K (B)^i \textit{i}_K (B)_i= \sigma^i \sigma_i.
\eea
 If $M$ is simply connected, using \eqref{dsigma=0}, we infer that there exists a function $\sigma: M \to \mathbb{C}$, called the Ernst potential, such that $\sigma_i = D_i \sigma$. Note also that $D_i g(K, K)= 2 B_{ij} K^j= -Re(\sigma_i)$. Hence we can choose the potential $\sigma$ such that $$Re(\sigma)=-g(K,K)=-X.$$ Therefore, equation \eqref{divergence-sigma} becomes, using \eqref{normsigma}:  
\bea
\label{wavesigma}
\Box_g \sigma = - X^{-1} D_i \sigma D^i \sigma + 2 R_{ij} K^i K^j
\eea

We define the contractions\footnote{In the case when the Killing vector field $K$ is $\partial_t$ those decompositions correspond to the electric and magnetic part of the electromagnetic tensor respectively.} of the electromagnetic tensor $F$ and its Hodge dual with the Killing vector field $K$ as follows:
\beaa 
E_i= \textit{i}_K (F)_i=K^j F_{ji}, \qquad H_i=\textit{i}_K (*F)_i=K^j *F_{ji}
\eeaa
These contractions uniquely determine the electromagnetic tensor $F$:
\bea
\label{decompositionF}
g(K,K) F_{ij}= K_i E_j - K_j E_i +\epsilon_{ijkl}  K^k H^l.
\eea
The Einstein-Maxwell equation implies
\bea 
R_{ij} K^i K^j = (2 F_{im} {F^m}_j - \frac 1 2 g_{ij} F^2)K^i K^j = 2 E^2 - \frac 1 2 X F^2
\eea
and since $F^2 = X^{-1} E^2$,  equation \eqref{wavesigma} becomes 
\bea\label{box-sigma}
\Box_g \sigma = - X^{-1} D_i \sigma D^i \sigma + 3 E^2
\eea
Writing $\sigma= - X - i Y$ we deduce from \eqref{box-sigma}
\bea
\square X&= X^{-1}(D^i X D_i X - D^i Y D_i Y) + 3 E^2 \label{waveeqXY}\\
\square Y&= 2X^{-1} D^i X D_i Y \label{waveeqY}
\eea

We derive now equations for the electric and the magnetic part of the electromagnetic tensor. Since $dE=d(\textit{i}_K F)= - \textit{i}_K (dF) + \mathcal{L}_K F=0$ by the Maxwell equations, if $M$ is simply connected there exists an electric potential $E: M \to \mathbb{C}$ such that $E_i= D_i E$. Similarly there exists a magnetic potential $H_i= D_i H$. The Maxwell equations then translate into wave equations for $E$ and $H$. We have 
\beaa
\Box_g E&=& D^i D_i E= D^i E_i = D^i (K^j F_{ji})= D^i K^j F_{ji}=- B^{ij} F_{ij}\\
\Box_g H&=& D^i D_i H= D^i H_i = D^i (K^j *F_{ji})= D^i K^j *F_{ji}=- B^{ij} *F_{ij}
\eeaa
We define the electromagnetic potential $\psi$ as $\psi=E+ i H$. The above equations then become
\bea\label{wave-psi-g}
\Box_g \psi= - B^{ij} (F_{ij} + i *F_{ij})= - B^{ij} \mathcal{F}_{ij}
\eea
where $ \mathcal{F}=F + i *F$.
Writing $B^{ij}$ in terms of $\sigma$ using (\ref{decompositionA}) and writing $\mathcal{F}_{ij}$ in terms of $E$ and $H$ using (\ref{decompositionF}), we get a closed system of equations for the Ernst potential and the electromagnetic potential.

\subsection{Induced equations on a hypersurface}

Let $N$ be a $2+1$ dimensional manifold, with an embedding of $N$ into $M$ transversal to the integral curves of $K$. The null second fundamental form of $K$ is given by $B$: 
\beaa
\RN{2}(X,Y)=g(D_X K, Y)=B(X,Y).
\eeaa
 We can introduce local coordinates $x^0, x^1, x^2, y$ in a neighborhood of a point $p \in N$ such that $K= \partial_{y}$. In these coordinates the metric $g$ takes the form 
 \beaa
 g= X(dy + A)^2 + h_{ij}dx^i dx^j \qquad \text{with $A=A_i dx^i$}
 \eeaa 
 We have that $X^{-1} K_i dx^i = dy + A_i dx^i$, therefore 
 \beaa
 dA=d\left(X^{-1} K_i dx^i\right)=X^{-1} D_j K_i dx^j \wedge dx^i - X^{-2} D_j \sigma K_i dx^i \wedge dx^j = X^{-1}\left(B_{ij} + X^{-1} K_i D_j \sigma\right) dx^i \wedge dx^j.
 \eeaa
  Using (\ref{decompositionB}), we have 
  \beaa
  dA=\frac 1 2 X^{-2}\left(K_i D_j \sigma + K_j D_i \sigma - \epsilon_{ijkl} K^k D^l Y\right) dx^i \wedge dx^j= - \frac 1 2 X^{-2}\left(\epsilon_{ijkl} K^k D^l Y\right) dx^i \wedge dx^j.
 \eeaa
 Writing the relation between the volume form of the 3-dimensional metric $h$ and the 4-dimensional $g$, i.e. ${\epsilon^{(h)}}_{ijk}= X^{-1/2}\epsilon_{ijkm} K^m$, and denoting $\nabla$ the covariant derivative with respect to $h$, we deduce 
 \beaa
 dA=\nabla_j A_i dx^j \wedge dx^i = \frac 1 2 X^{-3/2}({\epsilon^{(h)}}_{ijk} \nabla^k Y) dx^i \wedge dx^j.
 \eeaa
 We therefore obtain the following equation for $A$, valid for any Lorentzian 4-dimensional metric:
\beaa
\nabla_i A_j - \nabla_j A_i = X^{-3/2}({\epsilon^{(h)}}_{ijk} \nabla^k Y)
\eeaa

We now express equations \eqref{box-sigma}, \eqref{waveeqXY} and \eqref{wave-psi-g} in terms of the induced metric $h$. 

Since $K(\sigma)=0$, we deduce that $D_i \sigma= \nabla_i \sigma$. We obtain
\beaa 
\Box_h \sigma&=& \Box_g \sigma - X^{-1} K^i K^j D_i D_j \sigma =\Box_g \sigma - X^{-1}K^i D_i (K^j D_j \sigma)+ X^{-1}K^i D_i K^j D_j \sigma\\
&=&\Box_g \sigma + X^{-1} K^i {B_i}^j D_j \sigma= \Box_g \sigma - \frac 1 2 X^{-1} D^i X D_i \sigma.
\eeaa
The induced version of equations \eqref{box-sigma} and \eqref{waveeqXY} are therefore
\beaa
\Box_h \sigma&=& - X^{-1}(D^i \sigma D_i \sigma + \frac 1 2 D^i X D_i \sigma) + 3 E^2 \\
\Box_h X&=& X^{-1} (\frac 3 2 \nabla^i X \nabla_i X - \nabla^i Y \nabla_i Y) + 3 E^2 \\
\Box_h Y&=& \frac 5 2 X^{-1} \nabla^i X \nabla_i Y
\eeaa
These are the normal-normal and normal-tangential components of the Einstein-Maxwell equations. The tangential-tangential components can be deduced by the Gauss-Codazzi equation:
\beaa 
R_{ij} - X^{-1} R_{iKjK}= {R^{(h)}}_{ij} + X^{-1} {\RN{2}_j}^m \RN{2}_{im}.
\eeaa
 From the Killing equation, we deduce 
 \beaa
 R_{iKjK}=K^m D_i D_m K_j = D_i (K^m D_m K_j)=D_i \left(-\frac 1 2 D_j X\right)=-\frac 1 2 D_i D_j X= -\frac 1 2 \nabla_i \nabla_j X.
 \eeaa
  Thus the Gauss-Codazzi equation reduces to 
\bea
\label{Riccireduced}
{R^{(h)}}_{ij}= R_{ij} + \frac 1 2 X^{-1} \nabla_i \nabla_j X - X^{-1} {B_j}^m B_{im}
\eea

Since $K$ preserves the electromagnetic tensor, we have that $K(\phi)=0$. We therefore obtain the induced version of \eqref{wave-psi-g}: 
\beaa
\Box_h \psi = -B^{ij} \mathcal{F}_{ij} - X^{-1} D^i X D_i \psi
\eeaa
This equation for the electromagnetic potential completes the system of closed equations for the reduced Einstein-Maxwell equations.

\vspace{1cm}

\begin{flushleft}
\small{DEPARTMENT OF MATHEMATICS, COLUMBIA UNIVERSITY} \\
\textit{E-mail address}: \href{mailto:egiorgi@math.columbia.edu}{egiorgi@math.columbia.edu}
\end{flushleft}
\end{document}